\documentclass[prb, preprint, twocolumn, showpacs, amsmath, amssymb]{revtex4-1}

\usepackage{graphicx}
\usepackage{dcolumn}
\usepackage{bm}
\usepackage{multirow}
\usepackage{epstopdf}
\usepackage{textcomp, amsmath}

\usepackage{color}
\usepackage{graphicx}

\begin{document}
\title{Quantum Monte Carlo applied to solids}

\author{Luke Shulenburger}
\email{lshulen@sandia.gov}

\author{Thomas R. Mattsson}
\email{trmatts@sandia.gov}
\affiliation{Sandia National Laboratories, Albuquerque, New Mexico 87185, USA}

\date{\today}

\begin{abstract}
We apply diffusion quantum Monte Carlo (DMC) to a broad set of solids, benchmarking the method by comparing bulk structural properties (equilibrium volume and bulk modulus) to experiment and DFT based theories.  The test set includes materials with many different types of binding including ionic, metallic, covalent and van der Waals.   We show that, on average, the accuracy is comparable to or better than that of density functional theory (DFT) when using the new generation of functionals, including one hybrid functional and two dispersion corrected functionals.  The excellent performance of quantum Monte Carlo (QMC) on solids is promising for its application to heterogeneous systems and high-pressure/high density conditions.  Important to the results here is the application of a consistent procedure with regards to the several approximations that are made, such as finite-size corrections and pseudo-potential approximations.  This test set allows for any improvements in these methods to be judged in a systematic way.

\end{abstract}

\preprint{{\large SAND 2013-7693J}}

\maketitle

Although density functional theory (DFT) \cite{HohenbergKohn, KohnSham} is a remarkably successful theory for calculating many properties of matter, DFT does not necessarily constitute the endgame of predictive quantum mechanical methods. Increasing requirements for accuracy, ability to make systematic improvements, and predictability for a broader range of materials continue to drive research in fundamental electronic structure theory.  By solving the Schr{\"o}dinger equation directly using stochastic sampling methods, quantum Monte Carlo (QMC) and specifically diffusion Monte Carlo (DMC) offers an approach complementary to that of DFT.\cite{dmc-solids-review,ncstate-review,cambridge-review}  In addition, the impending shift in computational paradigm offers opportunities for methods that are inherently suited for heterogeneous architectures~\cite{gpu-qmc}. In this letter, we present a study of the performance of DMC for a wide set of solids: insulators, semiconductors, simple metals, and transition metals. We anticipate that the result of the study will serve as a guide for when to employ DMC for condensed systems as well as a foundation from which systematic improvements can be gauged.

New QMC codes\cite{CASINO,qmcpack,qwalk,champ} and ever increasing computer power are making it feasible to perform diffusion Monte Carlo calculations of a broader range of systems and problems, just as DFT over the course of the last twenty years moved from being a niche method practiced by a small number of experts to being a widely used technique.  In the last ten years, QMC calculations on energy barriers,\cite{qmc-energy-barrier} energy differences between solid phases,\cite{cohen-quartz-stishovite,GaAs-VMC} strongly correlated materials,\cite{mitas-FeO} van der Waals systems,\cite{cambridge-noble-gases} and interfaces between surfaces and molecules where van der Waals is blended with covalent and hydrogen bonding\cite{alfe-catalysis} have all appeared in the literature with impressive accuracy.  Several of these applications depend on the capability to calculate physics such as van der Waals interactions or the interplay between localization and itinerancy that are difficult to precisely include in semi-local Kohn-Sham density functional schemes.  Additionally, all of these applications require the calculation of energy differences significantly smaller than the previous holy grail of electronic structure methods: the 1 kcal/mol accuracy often described as chemical accuracy.

Despite these successes, there is still an open fundamental question regarding the accuracy of diffusion quantum Monte Carlo methods.  While the method is in principle exact, several approximations are introduced in practice.  These approximations range from fundamental considerations such as mitigation of the fermion sign problem to numerical issues such as the finite size of the time steps.  Due to the relatively large computational cost of QMC calculations, only rarely have calculations of multiple systems been presented in the same work.  Additionally, there have been numerous advances in the treatment the systematically controllable approximations inherent in QMC applied to solids and due to the plethora of choices available, only rarely have two calculations been performed using the exact same technique.  The present article aims to remedy this shortcoming by using qmcpack to perform DMC calculations of the energy versus volume for a wide variety of materials.  Using the present approach as a benchmark has numerous advantages in that it can test several different regimes of interaction while the end result of the series of calculations is two numbers: the equilibrium volume, and bulk modulus, both of which can be directly compared to experimental data with high accuracy as well as results from other theoretical methods.

The particular set of approximations assessed in this paper are selected from the state of the art in an effort to perform high accuracy calculations and also to minimize the dependence on the mean field method chosen to generate the trial wavefunctions and pseudopotentials.  There are however two notable exceptions.  The trial wavefunctions which are used for importance sampling and defining the nodal surface are of a Slater-Jastrow form with the single particle orbitals coming directly from the ground state of the DFT calculations.  No efforts have been made to improve them through for instance optimizing correlated wavefunctions.  Also, while great care was taken in the generation and application of pseudopotentials, these are developed within the mean field theory and no additional effort (such as performing auxiliary calculations on all electron systems\cite{esler-cBN}) is taken to optimize them for many body calculations.  Although both of these approximations are the subject of research, there exists as of yet no definitive solutions that can be applied at a reasonable cost in condensed systems.
{\em The set of approximations made in this work can together be characterized as the at present best possible while computationally realistic to perform for a large enough set of calculations to allow systematic conclusions to be made.} The approximations are thus representative of the level of theory available for successful application of QMC to a broad range of problems in material physics, high-pressure physics, catalysis, biophysics, and many other fields.

The accuracy of the DMC calculations are determined by three necessary approximations: approximating the behavior of a solid in the thermodynamic limit from calculations on a finite size simulation cell with periodic boundary conditions, the fixed node approximation,~\cite{fixed-node} and the use of pseudopotentials to obviate the need to calculate properties of chemically inert core electrons.  With finite computing time and state of the art methods it is not possible to reduce any of these to zero except in special cases, so in order to establish a baseline of accuracy for DMC, a consistent set of approximations must be chosen.  In this work, the following choices are made:  Firstly the one body-finite size effects are treated with twist averaged boundary conditions\cite{twist-averaging}, the MPC is used to mitigate the two body potential energy errors\cite{MPC,drummond-finite-size} and the Chiesa corrections using the Jastrow factors optimized within VMC are used to mitigate the kinetic contribution to the two body finite size effects.  Secondly, the instabilities resulting from the fermion sign problem are mitigated using the fixed node approximation where the nodes come from a Slater-Jastrow wavefunction with single particle wavefunctions coming from the Local Density Approximation (LDA) to Density Functional Theory.  Finally, pseudopotentials are constructed using the opium~\cite{opium} pseudopotential generation code using the LDA and checking to ensure that ghost states were not introduced when performing calculations using the Kleinmann-Bylander form.~\cite{kleinmann-bylander}.

A critical part of this study was the construction of pseudopotentials suitable for use in solid state DMC calculations.  Other pseudopotential sets \cite{trail-needs, filippi-dolg} were considered but eschewed due to two factors.  The first of these factors is the widespread use of a Kleinmann-Bylander~\cite{kleinmann-bylander} technique of representing the pseudopotentials used in most modern plane wave based DFT codes.  Given the desirability of recasting the semi-local pseudopotentials referenced above in a Kleinmann-Bylander form, a problem arises that ghost states are often introduced resulting in a Hamiltonian significantly different than intended, unless the pseudopotential has been constructed with this in mind.   The second of these factors concerns the fidelity of the Hamiltonian that results from the use of the pseudopotential approximation.  Ideally, at the single particle level, the relative eigenvalues of the valence states should not be shifted by the introduction of a pseudopotential.  It is common in the literature to note the inability of extant QMC minded pseudopotential sets to do this and subsequent ad hoc corrections are common.\cite{alfe-MgO,diamond-beta-sn,esler-cBN}   A first step in reducing the size of this approximation is to use pseudopotentials that have been validated versus all electron calculations in the same environment (I.e. in a condensed phase).  While this is still intractable within DMC for many materials, it is possible to perform such calculations within the DFT framework in which the pseudopotentials were produced.

\begin{table}[h]
\begin{tabular}{@{}lllll@{}}
\hline
 - & $V_0$ & $V_0^{AE}$ & $B_0$ & $ B_0^{AE}$ \\ \hline
Al   & 105.62  & 106.84 & 83.1 & 82.5 \\
Ar  & 208.0 & 204.4 & 6.71 & 6.83 \\
Be  & 49.95 & 50.03 & 125.5 & 124.3 \\ 
BN  & 39.33 & 40.43 & 395 & 400 \\ 
BP  & 76.04 & 76.612 & 170.1 & 174.0 \\ 
C   & 37.57 & 37.23 & 466 & 465 \\ 
Kr  & 260.4 & 264.4 & 6.17 & 6.00 \\ 
Li  & 128.25 & 128.22 & 15.08 & 15.00 \\ 
LiCl  & 104.11 & 103.31 & 40.8 & 41.0 \\ 
LiF  & 50.48 & 50.50 & 87.0 & 86.2 \\ 
SiC  & 68.43 & 68.81 & 226.9 & 226.0 \\ 
Si  & 132.2 & 132.2 & 96.0 & 95.4 \\ 
Xe  & 345.2 & 352.6 & 5.79 & 5.03 \\ 
\hline
\end{tabular}
\caption{Columns 2 and 4 contain the parameters for a Vinet equation of state fit to converged LDA calculations performed with the pseudopotentials used in this paper versus all electron calculations either performed using LMTO with RSPT\cite{AM05-solids} (for Al, BN, BP, C, Li, LiCl, LiF, SiC an Si) or LAPW calculations performed using elk\cite{elk} (for Ar, Kr, Xe and Be).  Equilibrium volumes are given in bohr$^3$ per formula unit and bulk modulus in GPa}
\label{pseudo-accuracy}
\end{table}

While it would be ideal to use a PAW construction~\cite{Blochl} to eliminate the core electrons due to its excellent transferability and ability to accurately treat the higher energy states / angular momentum channels that are exercised by correlated methods~\cite{hennig} this is currently not possible because of the lack of a suitable algorithm to treat the nonlocal projectors that are necessary for this formalism within DMC.  We have instead chosen to generate norm conserving pseudopotentials with Kleinmann-Bylander projectors and no nonlinear core corrections using the opium code \cite{opium} and taking advantage of the optimized pseudopotential construction of Rappe\cite{rappe-optimized-pseudopotentials}.  These are then tested against LAPW or LMTO calculations of the energy versus volume of representative solid phases of the elements in question.  A very high degree of agreement is demanded between the equilibrium lattice constant (volume) and bulk moduli of the pseudopotential and all electron calculations.  The discrepancies between these two methods is shown in table~\ref{pseudo-accuracy} and is typically less than 0.1\% in the lattice constant and 0.3\% in the bulk modulus.  Due to these requirements, the cutoff radii are typically small and the resulting pseudopotentials are rather hard (see Table~\ref{converged-parameters}), requiring a large number of plane waves to accurately represent them.  While this introduces additional demand on the DFT calculations used to generate trial wavefunctions, it does not affect the cost of the DMC calculations, which employ a real space b-spline representation to evaluate the trial wavefunction.\cite{b-spline-rep,einspline}  However, the effects of using such hard pseudopotentials on the size of the locality approximation is potentially larger than for softer potentials particularly if the potential is significantly different for the different angular momentum channels.  The suitability of these pseudopotentials for DMC will largely rest on the cancellation of these errors (largely confined to the core region) between calculations of similar phases.  Such pseudopotentials frequently result in dynamical instabilities in the walker population when treated within the locality approximation,\cite{locality-approximation} so Casula t-moves are used throughout this work.\cite{casula-t-moves}

With the Hamiltonian fixed by the choice of pseudopotentials, the task is to compute the equation of state (energy versus volume) for the most stable phase of a wide range of materials and determine the accuracy of the equilibrium properties (equilibrium volume and bulk modulus) compared to experiment.  These calculations are performed using a Slater-Jastrow form of the trial wavefunction with one and two body Jastrow factors.  While other wavefunctions (such as inhomogeneous backflow\cite{backflow} or multideterminant expansions\cite{multideterminant}) could possibly result in improved nodal surfaces, these are beyond the scope of the current paper which intends to probe the accuracy of the most common wavefunctions as evidenced by their extensive use in the literature.\cite{dmc-solids-review,ncstate-review,cambridge-review}

\begin{figure}
\includegraphics[width=2.4in,angle=-90]{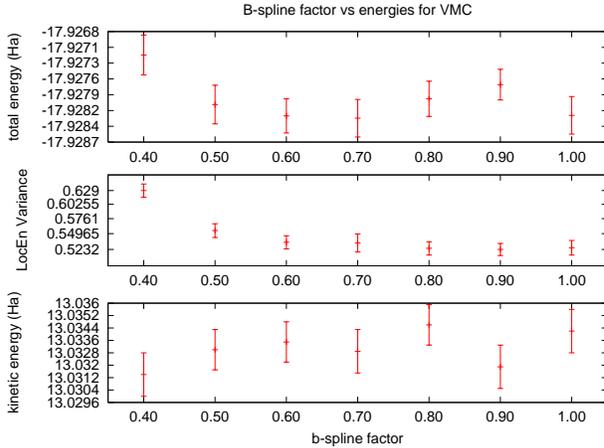}
\caption{(color online) Three panels showing the convergence of the total energy, the variance and the kinetic energy as the spacing between the b-splines representing the Slater determinant part of the wavefunction is decreased.  Larger spline factors correspond to having more real space basis functions.  In fact, the memory used to store the wavefunction goes as the factor on the x axis cubed.  In this case the total energy and kinetic energy have converged by a spline spacing of 0.5 but the variance still appears to be decreasing until 0.6.  In this way the memory necessary to perform the calculations can be tuned to both give accurate answers and if possible optimal variances within VMC.}
\label{spline-spacing-figure}
\end{figure}

The procedure for performing the quantum Monte Carlo calculations proceeds as follows.  DFT calculations are performed for the primitive cell of the material under construction both near ambient conditions and at elevated pressure (typically 300 GPa).  Then the Slater determinant part of trial wavefunctions are extracted using the converged charge density and k-points corresponding to the supercell and twisted boundary conditions under consideration.  Supercells of 16 to 32 atoms are typically used to avoid overly large finite size effects from biasing the convergence during this initial stage of calculation.  The Jastrow factors are optimized using a VMC calculation for purely periodic boundary conditions and these optimized Jastrow factors are used for all twists.  For these optimization calculations, the b-spline spacing is set equal to twice the resolution of the plane waves needed to converge the stress tensor in the DFT calculation.  Next, using this converged Jastrow factor the total energy and variance of a VMC calculation are converged with respect to the density of the b-splines.  This step is necessary to avoid exhausting the memory available for individual nodes of the computer on which the calculations are performed for larger supercells.  Typically the energy converges much faster than the variance (shown in Fig.~\ref{spline-spacing-figure}) although when possible, using the finer mesh needed to converge the variance is worthwhile as it decreases the computational time necessary for the much more expensive DMC calculations.

\begin{figure}
\includegraphics[width=2.4in,angle=-90]{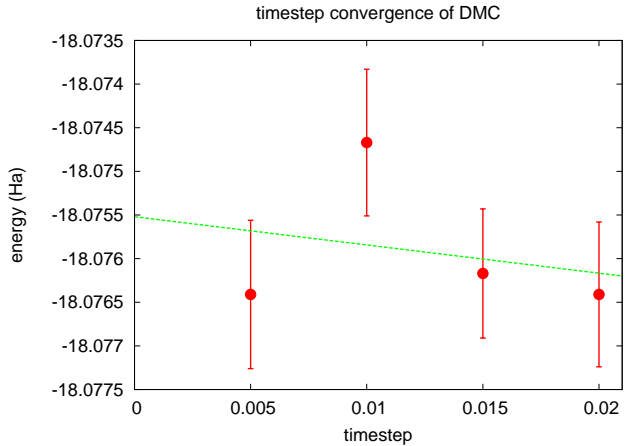}
\caption{(color online) Representative calculation of convergence of DMC energy with respect to timestep.  In all cases the energy is converged to within 1 mHa per atom.  For most calculations, the timestep error is smaller than the statistical error}
\label{timestep-convergence}
\end{figure}

\begin{figure}
\includegraphics[width=2.4in,angle=-90]{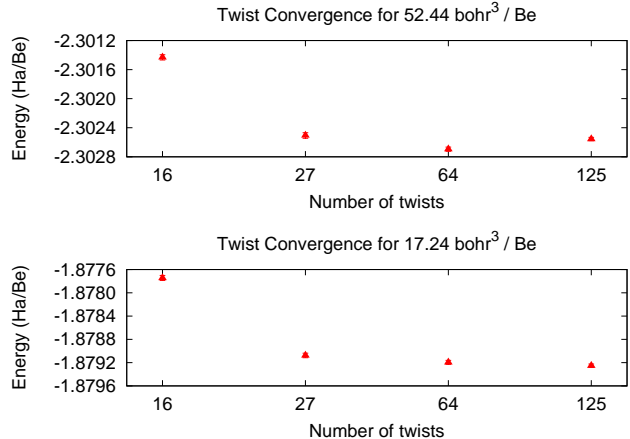}
\caption{(color online) Representative calculations of convergence of DMC energy with respect to twist averaging.  The top panel shows the energy per beryllium atom for a volume near the ambient pressure.  The bottom panel shows the energy for the same system, but at a pressure near 300 GPa.  In both cases the energy has converged to well within 1 mHa per atom for 64 supercell twists.}
\label{twist-convergence}
\end{figure}

Next DMC calculations are performed for these same moderate sized supercells of the material in question in order to converge the DMC timestep and the number of supercell twists necessary to reduce one body finite size effects.  These calculations are performed both for the material near solid density and also at a pressure corresponding roughly to 300 GPa in the LDA so as to account for the difference between the electronic structure between the ambient and high pressure material.  The timestep is taken to be converged whenever the energy is within 1 mHa per atom of the extrapolation to zero timestep (shown in Fig.~\ref{timestep-convergence}).  The one body finite size effects are similarly converged to the 1 mHa per atom level by simultaneously increasing the density of the grid of supercell twists in reciprocal space in each direction, this is shown in Fig.~\ref{twist-convergence}.  

Furthermore, the two body finite size effects are studied by using a succession of supercell sizes generated in an effort to maximize the simulation cell radius given the number of atoms in the supercell.  Supercells are chosen by finding the arbitrary tiling of the primitive cell such that the size of an inscribed sphere is maximized.  This construction is selected to maximize the distance between periodic images of electrons and thus decrease spurious correlations.  Unfortunately, it was not possible to achieve absolute convergence to within the desired 1 mHa / formula unit.  However, this level of convergence is not a necessary condition for achieving converged values for the equilibrium volume and bulk modulus, properties which depend only upon relative energies between different densities of the material.  To this end, calculations were performed for different sized supercells for both the ambient and 300 GPa densities.  Calculations were performed with supercell twists given by the earlier convergence tests and taking into account the two body finite size corrections described earlier.  Convergence was deemed to have been achieved when the change in total energy from one supercell size to the next is equal within 0.1 mHa / formula unit.  As a rule, the total energy was also converged for each volume to within 2 mHa although this was not universally the case.  The final parameters used for each material are summarized in Table~\ref{converged-parameters}.

\begin{table}[h]
\begin{tabular}{@{}cccccc@{}}
\hline
 - & DFT $E_{cut}$ & spline factor & timestep & twists & supercell size \\ \hline
Al  & 75 & 0.65  & 0.01 & 64 & 108 \\
Ar & 110 & 1.0 & 0.01 & 8 & 108 \\
Be & 120 & 0.8 & 0.01 & 27 & 66 \\ 
BN & 100 & 0.65 & 0.005 & 64 & 32 \\ 
BP & 112.5 & 0.75 & 0.005 & 64 & 32 \\ 
C  & 105 & 0.8 & 0.0025 & 64 & 32 \\ 
Kr & 110 & 1.0 & 0.01 & 8 & 108 \\ 
Li & 225 & 1.0 & 0.005 & 216 & 28 \\ 
LiCl & 212.5 & 1.0 & 0.01 & 27 & 32 \\ 
LiF & 212.5 & 1.0 & 0.005 & 27 & 32 \\ 
SiC & 125 & 0.75 & 0.005 & 27 & 32 \\ 
Si & 75 & 0.7 & 0.005 & 125 & 32 \\ 
Xe & 125 & 0.8 & 0.005 & 8 & 108 \\ 
\hline
\end{tabular}
\caption{Converged parameters for the DMC calculation reported in this work.  The DFT cutoff energies are given in Hartree and together with the spline factor define the basis in which the QMC wavefunctions are represented.  The supercell sizes are multiples of the smallest primitive cell that could be constructed for the given symmetry.  In the case of the monatomic systems, this is 1 atom per cell except for Si and C which require 2.  Likewise, all of the biatomic systems have 2 atoms in the primitive cell.}
\label{converged-parameters}
\end{table}

Finally, using the converged parameters for these simulations, we performed a series of calculations at a variety of lattice constants equally spaced by 3\% of the equilibrium lattice constant and ranging from -10 GPa to 300 GPa according to the LDA results.  the resulting energy vs volume curves are fit using the Vinet form of the equation of state\cite{vinet} to extract the equilibrium volume and bulk modulus.  As expected, these quantities show sensitivity to the range of E(V) data which is included in the fit: in all cases we have reported the fit to the full range from the largest volume calculated to the smallest, corresponding roughly to a pressure range of -10 to 300 GPa.  In reporting these quantities, we show the errors resulting from the least square fit of our results to the Vinet form.

\begin{figure}
\includegraphics[width=2.4in,angle=-90]{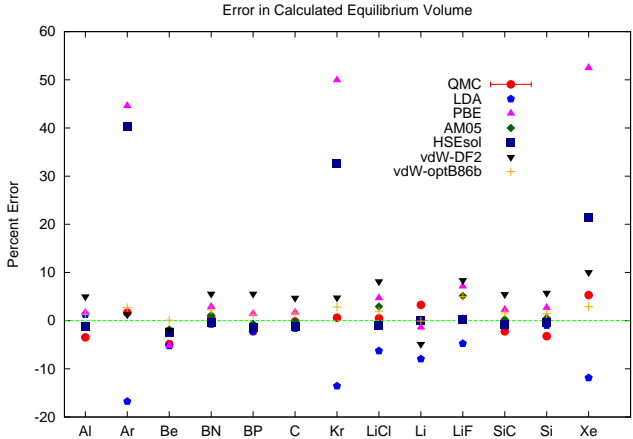}
\caption{(color online) Percentage error in the calculated equilibrium volumes from each of the different theories as compared to experiment.}
\label{volume-error-figure}
\end{figure}
\begin{figure}
\includegraphics[width=2.4in,angle=-90]{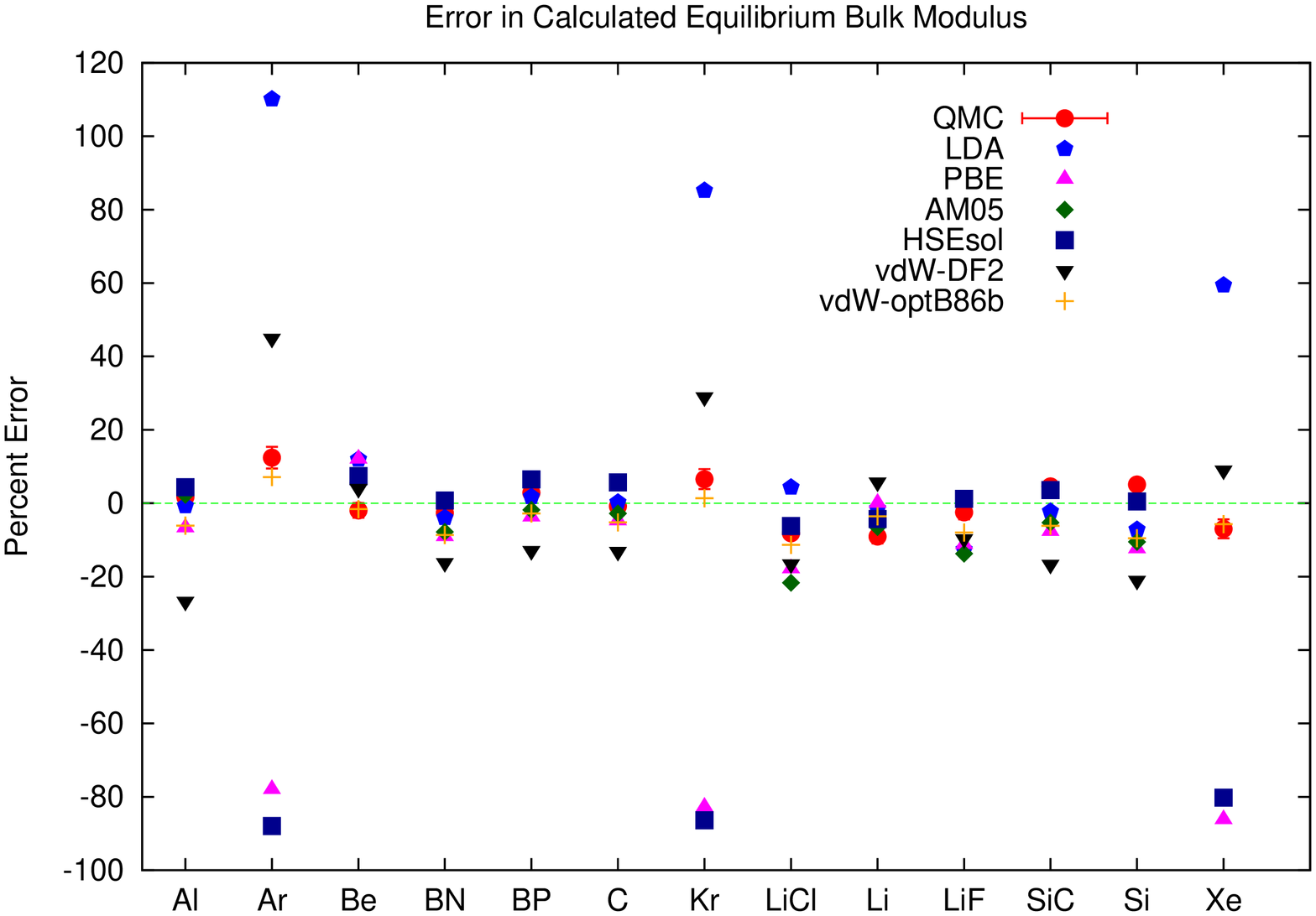}
\caption{(color online) Percentage error in the calculated equilibrium bulk modulus from each of the different theories as compared to experiment.}
\label{bulk-modulus-error-figure}
\end{figure}

When faced with a research challenge that requires first-principles simulations, a major consideration when choosing approach is the performance of a method, for example QMC, in relationship to other methods. In addition to the results from QMC, we therefore report seven other sets of results using the same protocol for calculating the lattice constant and bulk moduli. Firstly, we compare to DFT with the LDA functional.\cite{ceperley-alder,perdew-zunger}  Also, we compare results with the most popular GGA functional, PBE\cite{PBE} and a next generation GGA functional, AM05.\cite{AM05}  We also include calculations using a state of the art hybrid functional, HSEsol, which has been shown to deliver excellent results for these sort of calculations\cite{HSE-sol} but is computationally much more expensive than semi-local DFT.  Given the difficulties of these standard functionals in treating van der Waals systems like the noble gas solids in this test set, we also include two functionals designed for this purpose, vdW-DF2\cite{vdw-DF2} and vdW-optB86b\cite{vdw-optB86b}.  Finally, we compare our results to experiments, corrected for zero point motion and thermal expansion where appropriate as detailed in work by Schimka et al.\cite{HSE-sol}  The results for the equilibrium volume are presented in table \ref{volume-results-table} and also graphically in figure \ref{volume-error-figure}.  Likewise, the results for the bulk modulus are detailed in table \ref{bulk-modulus-results-table} and shown graphically in figure \ref{bulk-modulus-error-figure}.

\begin{table*}[h]
\begin{tabular}{@{}cccccccccc@{}}
\hline
\text{material}  &  \text{DMC}  &  \text{statistical error}  &  \text{LDA}  &  \text{PBE}  &  \text{AM05} &  \text{HSEsol}  &  \text{vdW-DF2}  &  \text{vdW-optB86b} & \text{experiment} \\
\hline
\hline
Al  &  105.650  &  0.067  &  110.832   &  111.245  &  108.298 &  108.217    &  114.920   &  110.915 &  110.585 \\
Ar  &  248.352  &  1.224  &  203.383   &  353.252  &      -   &  342.882    &  247.494   &  250.907 &  252.805 \\
Be  &  51.140    &  0.143 &  50.956    &  50.956   &  52.766  &  52.401     &  52.735    &  53.810  &  54.776  \\
BN  &  78.796   &  0.024  &  77.603    &  80.430   &  79.041  &  77.863     &  82.549    &  79.928  &  79.173  \\
BP  &  152.844  &  0.130  &  152.815   &  158.603  &  155.281 &  154.146    &  165.018   &  158.348 &  157.663 \\
C   &  37.762    &  0.042 &  37.231    &  38.477   &  37.771  &  37.358     &  39.619    &  38.445  &  38.284  \\
Kr  &  299.386  &  1.566  &  257.230   &  446.206  &       -  &  394.782    &  311.798   &  306.055 &  303.646 \\
LiCl&  220.900  &  0.297  &  206.114   &  230.172  &  226.304 &  217.534    &  237.766   &  224.189 &  224.584 \\
Li   &  143.455  &  0.302 &  127.878   &  136.995  &  139.159 &  138.917    &  132.151   &  138.797 &  141.834 \\
LiF  &  106.096  &  0.212 &  100.693   &  113.240  &  111.163 &  105.882    &  114.582   &  110.998 &  108.785 \\
SiC &  135.400  &  0.026  &  136.962   &  141.665  &  138.869 &  137.342    &  146.077   &  140.696 &  139.636 \\
Si   &  130.062  &  0.050 &  133.049   &  137.985  &  135.128 &  133.938    &  142.112   &  136.326 &  135.054 \\
Xe  &  404.780  &  1.275  &  338.758   &  586.105  &       -  &  466.665    &  423.048   &  395.488 &  388.952 \\
\hline
ME               &  -1.08      & 0.05  &  -3.88  &  2.69  & 1.09  & -0.93      & 5.46 & 1.96 &  - \\
MAE              &   2.39      & 0.05  &   4.16  &  3.63  & 1.73  &  0.96      & 7.02 & 1.98 &  - \\
MRE (\%)         &  -1.13      & 0.04  &  -2.96  &  1.73  & 0.65  & -0.88      & 4.18 & 1.65 &  - \\
MARE (\%)        &   2.10      & 0.04  &   3.21  &  3.10  & 1.39  &  0.91      & 5.53 & 1.66 &  - \\
\hline                                                                                            
ME (all)         &   1.18      & 0.19  & -12.75  & 37.39  &  -    & 20.67      & 8.51 & 3.51 &  - \\
MAE (all)        &   3.83      & 0.19  &  12.96  & 38.11  &  -    & 22.12      & 9.70 & 3.53 &  - \\
MRE (all) (\%)   &  -0.29      & 0.07  &  -5.52  & 12.68  &  -    &  6.59      & 4.46 & 1.92 &  - \\
MARE (all) (\%)  &   2.20      & 0.07  &   5.72  & 13.69  &  -    &  7.96      & 5.50 & 1.93 &  - \\
\hline
\hline
\end{tabular}
\caption{Results for the equilibrium volume of the solids as determined by a fit of the Vinet equation to calculations.  All values are given in $bohr^3$ per formula unit.  The experimental numbers have finite temperature thermal expansion and zero point energy subtracted following the work of Schimka et al\cite{HSE-sol}.  The DMC results include an error estimate due to the statistical error in the individual calculations.  The error statistics are calculated first excluding the noble gases and secondly including for all of the materials.  The four statistics compare the calculations to the experimental value and are the mean error (ME: $\sum x_{calc} - x_{expt}$), mean absolute error (MAE: $\sum |x_{calc} - x_{expt}|$), mean relative error (MRE: $\sum \frac{x_{calc} - x_{expt}}{x_{expt}}*100$) and mean absolute relative error (MARE: $\sum \frac{|x_{calc} - x_{expt}|}{x_{expt}}*100$).  Note that results are omitted for the AM05 functional as applied to the noble gases as it fails to bind by design and is thus not applicable.}
\label{volume-results-table}
\end{table*}

\begin{table*}[h]
\begin{tabular}{@{}cccccccccc@{}}
\hline
\text{material}  &  \text{DMC}  &  \text{statistical error}  &  \text{LDA}  &  \text{PBE}  &  \text{AM05} &  \text{HSEsol}  &  \text{vdW-DF2}  &  \text{vdW-optB86b} & \text{experiment} \\
\hline
\hline
        Al   &  83.35 &   0.58 &  81.40 &  76.50 &  83.90 &  85.60 &  60.10 &  77.00 &  82.00 \\ 
        Ar   &   3.80 &   0.10 &   7.10 &   0.74 &   0.00 &   0.41 &   4.90 &   3.62 &   3.38 \\ 
        Be   & 119.28 &   2.42 & 136.26 & 136.30 & 128.50 & 130.60 & 126.50 & 119.70 & 121.65 \\ 
        BN   & 399.34 &   1.92 & 394.00 & 373.00 & 378.00 & 413.30 & 343.80 & 374.70 & 410.20 \\ 
        BP   & 172.85 &   2.08 & 171.00 & 161.70 & 165.00 & 178.90 & 146.32 & 163.30 & 168.00 \\ 
         C   & 450.86 &   3.54 & 456.00 & 433.10 & 442.00 & 480.40 & 395.00 & 431.00 & 454.70 \\ 
        Kr   &   3.90 &   0.10 &   6.78 &   0.63 &   0.00 &   0.50 &   4.72 &   3.71 &   3.66 \\ 
      LiCl   &  35.53 &   0.48 &  40.40 &  31.80 &  30.30 &  36.30 &  32.30 &  34.30 &  38.70 \\ 
        Li   &  12.64 &   0.26 &  13.70 &  13.90 &  13.00 &  13.30 &  14.70 &  13.40 &  13.90 \\ 
       LiF   &  74.40 &   1.47 &  66.70 &  67.70 &  65.80 &  77.20 &  68.90 &  70.20 &  76.30 \\ 
       SiC   & 239.61 &   0.48 & 224.00 & 211.50 & 217.00 & 237.30 & 191.00 & 215.00 & 229.10 \\ 
        Si   & 105.95 &   0.44 &  93.60 &  88.30 &  90.20 & 101.30 &  79.60 &  91.20 & 100.80 \\ 
        Xe   &   3.60 &   0.10 &   6.17 &   0.53 &   0.00 &   0.77 &   4.22 &   3.65 &   3.87 \\ 
\hline
ME               &  -0.15      & 0.54  &  -1.83  &  -10.16  & -8.17  &  5.89      & -23.71 & -10.56 &  - \\
MAE              &   4.53      & 0.54  &   5.95  &   13.09  &  9.92  &  6.49      &  24.84 &  10.56 &  - \\
MRE (\%)         &  -1.10      & 0.40  &  -0.96  &   -6.15  & -6.22  &  1.94      & -12.31 &  -6.30 &  - \\
MARE (\%)        &   3.94      & 0.40  &   4.65  &    8.55  &  7.81  &  4.04      &  14.26 &   6.30 &  - \\
\hline                                                                                            
ME (all)         &  -0.09      & 0.42  &  -0.70  &   -8.54  &  -     &  3.82      & -18.01 &  -8.11 &  - \\
MAE (all)        &   3.55      & 0.42  &   5.28  &   10.76  &  -     &  5.70      &  19.34 &   8.16 &  - \\
MRE (all) (\%)   &   0.08      & 0.48  &  18.86  &  -23.73  &  -     &-18.09      &  -3.08 &  -4.63 &  - \\
MARE (all) (\%)  &   5.03      & 0.48  &  23.18  &   25.58  &  -     & 22.69      &  17.35 &   5.94 &  - \\
\hline
\hline
\end{tabular}
\caption{Results for the equilibrium bulk-modulus of the solids as determined by a fit of the Vinet equation to calculations.  All values are given in GPa.  The experimental numbers have finite temperature thermal expansion and zero point energy subtracted following the work of Schimka et al\cite{HSE-sol}.  The DMC results include an error estimate due to the statistical error in the individual calculations.  The error statistics are calculated first excluding the noble gases and secondly including for all of the materials.    The four statistics compare the calculations to the experimental value and are the mean error (ME: $\sum x_{calc} - x_{expt}$), mean absolute error (MAE: $\sum |x_{calc} - x_{expt}|$), mean relative error (MRE: $\sum \frac{x_{calc} - x_{expt}}{x_{expt}}*100$) and mean absolute relative error (MARE: $\sum \frac{|x_{calc} - x_{expt}|}{x_{expt}}*100$). Note that results are omitted for the AM05 functional as applied to the noble gases as it fails to bind by design and is thus not applicable.}
\label{bulk-modulus-results-table}
\end{table*}

At first, ignoring the noble gas solids, the DMC results tend to provide roughly equal fidelity compared to the experiments as HSEsol, with the absolute errors from the DMC tended to be slightly larger than those from HSEsol, but had a slightly smaller bias. The situation changes somewhat when the noble gas solids are included.  In this case, the HSEsol errors are considerably larger despite its explicitly nonlocal construction.  For this reason we considered the van der Waals functional vdW-DF2.  This functional had encouraging results for the noble gas solids, but performed poorly for the other materials.  Increasing the test set to include vdW-optB86b provided results which were on par with the accuracy of AM05, but were consistent when including the noble gases.

The results presented in Tables~\ref{volume-results-table} and \ref{bulk-modulus-results-table} validate the use of DMC for a broad range of solids by showing little bias across the test set while maintaining an overall high accuracy.  The outcome is for a number of reasons very promising for QMC.  First, the nodal surface employed by the fixed node DMC calculations in this study remains at the ground state DFT level; this implies that the sensitivity of the structural properties upon the nodal surface is not extraordinarily large.  The second encouraging observation is that the accuracy is not unduly influenced by the physics responsible for the chemical bonding.  Covalent, metallic and van der Waals solids are all described with roughly the same fidelity.  The behavior is fortunate given the desire to have a method which works well where semilocal DFT fails qualitatively, such as is the case for so called strongly correlated materials such as transition metal oxides.  Finally, it was possible to achieve these results with a consistent procedure and without making any corrections do to the use of pseudopotentials.  Therefore, we conclude that DMC will be an extraordinarily interesting technique going forward due to its accuracy and its extremely favorable parallel scaling.

All of this is not to say that this work represents the ultimate accuracy that is possible for the calculation of condensed matter with DMC.  In fact, there are several improvements to this methodology that would be interesting to explore going forward and benchmark to the present work. The first and perhaps most important question would be to understand the size of the pseudopotential approximation compared to the fixed node approximation.  Reducing the pseudopotential approximation's impact could be tried from several perspectives.  However, the most important of these should be attempting to perform all-electron calculations for at first light elements followed by progressively heavier ones.  Doing so would provide a baseline against which pseudopotential calculations can be compared.  The pioneering work of Esler et al\cite{esler-cBN} provides a first step along this path, performing highly accurate all electron calculations for cubic boron nitride.

However, it is not trivial to decompose the error in these calculations into fixed node and pseudopotential errors.  In the cubic boron nitride work, the subsequent agreement with experiment suggested that the fixed node error was also quite small, however the fixed node error may reasonably be expected to scale with the total energy of the calculation performed, that is when higher energy core states are included the fixed node error should get larger.  There is reason to believe that these errors may cancel for materials at different densities, although the same could be said of pseudopotential errors as well.  Ideally, such work would also include efforts to reduce the fixed node approximation as well.
Reducing the size of the fixed node approximation may be somewhat more difficult as existing advanced forms for improved wavefunctions are either extremely expensive to apply for large systems (such as backflow transformations) or pose additional difficulties for extended systems such as multideterminant expansions or geminals / pfaffians.
Finally, the finite size convergence of these and other QMC calculations remains a computational challenge. The unfavorable scaling of the computational cost of DMC with electron number remains a significant hurdle, however increasing levels of parallelism\cite{gpu-qmc} and larger computational facilities may allow for simulations of simple solids containing enough atoms to render this a small concern in the near future.

This article has provided a benchmark for structural properties of solids with DMC, however absolute energetics are also important in application and their calculation is a natural extension of this work.  Excellent results for atomization energies of small molecules have been presented by Morales et al.\cite{multideterminant}  For solids, defect energy calculations have featured in the literature, with results for MgO\cite{alfe-defect-MgO}, diamond\cite{hood-diamond} and aluminum\cite{hood-aluminum}.  Unfortunately, constructing a benchmark set of defect calculations may be difficult as the specifics of the crystal structure are important.  A perhaps simpler test would be calculation of the enthalpy of formation of various solids.  It would be useful in this case to change the methodology to provide a better description of the isolated atoms as the description of isolated atoms with a single Slater determinant times a Jastrow factor are likely worse than that of a solid.  This work would also require further convergence of finite size effects for the bulk as relative energetics would no longer be sufficient.  The potential impact of such a work is not to be understated however as this continues to be a particularly difficult test for density functional theory methods.

To conclude, in this article we have demonstrated that DMC today offers an increasingly attractive complement to DFT also for condensed matter systems by offering a different set of approximations, an opportunity for systematic increase in accuracy of the calculations, and most importantly a method with minimal bias between different types of binding: covalent, metallic, van-der Waals, and ionic.  We anticipate the results of this work to encourage a broader application of QMC to problems across many disciplines, including high-pressure science, physical chemistry, and material science.  

We thank Jeongnim Kim, Mike Desjarlais, Miguel Morales and Paul Kent for stimulating discussions and the Cielo Capability Computing Campaign for computer time. The work was supported by the NNSA Science Campaigns and LNS was supported through the Predictive Theory and Modeling for Materials and Chemical Science program by the Basic Energy Science (BES), Department of Energy (DOE).
Sandia is a multiprogram laboratory operated by Sandia Corporation, a Lockheed Martin Company, for the U.S. Department of Energy's National Nuclear Security Administration under Contract No. DE-AC04-94AL85000.

%

\bibliographystyle{prsty}
\bibliography{QMC_for_solids}



\end{document}